\documentclass[10pt,twocolumn]{wlscirep}
\usepackage{mystyle}
\graphicspath{{img/}}
\usepackage{siunitx}

\title{All-passive pixel super-resolution of time-stretch imaging}

\author[1]{Antony C.~S.~Chan}
\author[1]{Ho-Cheung Ng}
\author[1,2]{Sharat C.~V.~Bogaraju}
\author[1]{Hayden K.~H.~So}
\author[1]{Edmund Y.~Lam}
\author[1,*]{Kevin K.~Tsia}

\affil[1]{Department of Electrical and Electronic Engineering, the University of Hong Kong, Pokfulam, Hong Kong}
\affil[2]{Current address: Department of Computer Science and Engineering, National Institute of Technology Goa, India}
\affil[*]{tsia@hku.hk}

\begin{abstract}
Based on image encoding in a serial-temporal format,
optical time-stretch imaging entails a stringent requirement of state-of-the-art  fast data acquisition unit in order to preserve high image resolution at an ultrahigh frame rate --- hampering the widespread utilities of such technology. 
Here, we propose a pixel super-resolution (pixel-SR) technique tailored for time-stretch imaging that preserves pixel resolution at a relaxed sampling rate.
It harnesses the subpixel shifts between image frames inherently introduced by asynchronous digital sampling of the continuous time-stretch imaging process.
Precise pixel registration is thus accomplished without any active opto-mechanical subpixel-shift control or other additional hardware.
Here, we present the experimental pixel-SR image reconstruction pipeline that restores high-resolution time-stretch images of microparticles and biological cells (phytoplankton) at a relaxed sampling rate ($\approx 2\text{--}\SI{5}{\giga Sa\per\second}$) --- more than four times lower than the originally required readout rate ($\SI{20}{\giga Sa\per\second}$) ---
is thus effective for high-throughput label-free, morphology-based cellular classification down to single-cell precision.
Upon integration with the high-throughput image processing technology,
this pixel-SR time-stretch imaging technique represents a cost-effective and practical solution for large scale cell-based phenotypic screening in biomedical diagnosis and machine vision for quality control in manufacturing.
\end{abstract}

\begin{document}
\maketitle
\thispagestyle{fancy}
 
\section{Introduction}

High-speed optical imaging with the temporal resolution reaching the nanosecond or even picosecond regime is a potent tool to unravel ultrafast dynamical processes studied in a wide range of disciplines~\cite{Goda2009,Goda2013,Lau2016,Lei2016,Lau2016a}.
Among all techniques, optical time-stretch imaging not only can achieve an ultrafast imaging rate of MHz--GHz,
but also allow continuous operation in real time.
This combined feature makes it unique for ultrahigh-throughput monitoring and screening applications,
ranging from barcode recognition and web-inspection in industrial manufacturing~\cite{Yazaki2014} to imaging cytometry in life sciences and clinical diagnosis~\cite{Chen2016}.
Nevertheless, a key challenge of time-stretch imaging limiting its widespread utility is that the spatial resolution is very often compromised at the ultrafast imaging rate.
This constraint stems from its image encoding principle that relies on real-time wavelength-to-time conversion of spectrally-encoded waveform, through group velocity dispersion (GVD), to capture image with a single-pixel photodetector.
In order to guarantee high spatial resolution that is ultimately determined by the diffraction limit,
two interrelated features have to be considered.
First, sufficiently high GVD in a dispersive medium
($\approx \SI{1}{\nano\second\per\nano\meter}$ at the wavelengths of $1\text{--}\SI{1.5}{\micro\meter}$)
is needed to ensure the time-stretched waveform to be the replica of the image-encoded spectrum.
Second, time-stretch imaging inevitably requires the electronic digitizer with an ultrahigh sampling rate ($>\SI{40}{\giga Sa\per\second}$) in order to resolve the time-stretched waveform.
To avoid using these state-of-the-art digitizers, which incur prohibitively high cost, the common strategy is to further stretch the spectrally-encoded waveform with an even higher GVD such that the encoded image can be resolved by the cost-effective, lower-bandwidth digitizers. 
However, as governed by the Kramers-Kronig relations, high GVD comes at the expense of high optical attenuation that deteriorates the signal-to-noise ratio (SNR) of the images~\cite{Tsia2010}.
Although optical amplification can mitigate the dispersive loss,
progressively higher amplifier gain results in excessive amplifier noise,
which in turn degrades the SNR.
To combat against the nonlinear signal distortion and amplifier noise,
it also necessitates careful designs of multiple and cascaded amplifiers that complicate the system architecture.
Even worse, achieving high GVD-to-loss ratio becomes increasingly difficult as the operation wavelengths move from the telecommunication band to the shorter-wavelength window,
which is favourable for biomedical applications,
not to mention the benefit of higher diffraction-limited resolution at the shorter wavelengths.
This technical constraint of GVD explains that the overall space-to-time conversion achieved in time-stretch imaging is generally limited to few tens of picoseconds (or less) per resolvable image point.
As a consequence, it is common that the sampling rate of the digitizer,
i.e.\ the effective spatial pixel size,
is the limiting factor of the spatial resolution in time-stretch imaging.
In other words, the time-stretch image is easily affected by aliasing if sampled at a lower rate.

To address this challenge, we demonstrate a pixel super-resolution (pixel-SR) technique for enhancing the time-stretch image resolution while maintaining the ultrafast imaging rate.
It is possible because high-resolution (HR) image information can be restored from multiple subpixel-shifted, low-resolution (LR) time-stretch images captured by a lower sampling rate. 
Previously, we demonstrated that subpixel-shifted time-stretch image signal can be recorded in real time by pulse-synchronized beam deflection with the acousto-optic beam deflector (AOD)~\cite{Chan2015}.
However, it requires sophisticated synchronization control for precise sub-pixel registration at an ultrafast rate. 
It is also viable to perform subpixel-shifted time-stretch image capture by time-interleaving multiple commercial-grade digitizers (TiADC)~\cite{El-Chammas2012,Xu2016}.
Despite this, this approach is prone to inter-channel timing and attenuation mismatch errors, which degrade the system dynamic range and SNR.
It has also been demonstrated that the sampling rate can be effectively doubled by optical replication of spectrally-encoded pulses at a precise time delay~\cite{Dai2016}.
This approach, however, requires high-end test equipment for timing calibration, and is not easily scalable to achieve high resolution gain.

In view of these limitations of the existing techniques,
it is thus of great value if a \emph{passive} subpixel-shift scheme using a single commercial-grade digitizer at a lower sampling rate of $1\text{--}\SI{10}{\giga Sa\per\second}$ can be realized for time-stretch imaging.
Here, we propose a simple strategy to allow time-interleaved measurements by the inherent sampling clock drifting because
the digitizer sampling clock is unlocked from the pulse repetition frequency of the pulsed laser source.
By harnessing this effect at a lower sampling rate,
we are able to extract multiple LR time-stretch line-scans,
each of which is subpixel-shifted at the precision of tens of picoseconds.
This technique resembles the concept of equivalent time sampling adopted in high-end sampling oscilloscope~\cite{KeysightTechnologies2015,Tektronix2001}. 
In this paper, we demonstrate that the pixel-SR technique is able to reconstruct the HR time-stretch images at an equivalent sampling rate of $\SI{20}{\giga Sa\per\second}$,
from the LR images captured at $\SI{5}{\giga Sa\per\second}$.
In the context of imaging flow cytometry applications, we also demonstrate that pixel-SR facilitates the morphological classification of biological cells (phytoplankton).
Unlike any classical pixel-SR imaging techniques,
our method does not require any additional hardware for controlled subpixel-shift motion
(e.g.\ detector translation~\cite{Ben-Ezra2005,Shin2007}
illumination beam steering~\cite{Tkaczyk2012,Greenbaum2013}),
or complex image pixel registration algorithms for uncontrolled motions~\cite{Elad1997,Park2003},
thanks to the highly precise pixel drifting.
Therefore, this pixel-SR technique is in principle applicable to all variants of time-stretch imaging systems,
including quantitative phase time-stretch imaging~\cite{Lau2014,Mahjoubfar2013,Feng2016}.

\section{General concepts}
\label{sec:working-principle}

\begin{figure*}[tb]
\centering\includegraphics[width=\textwidth]{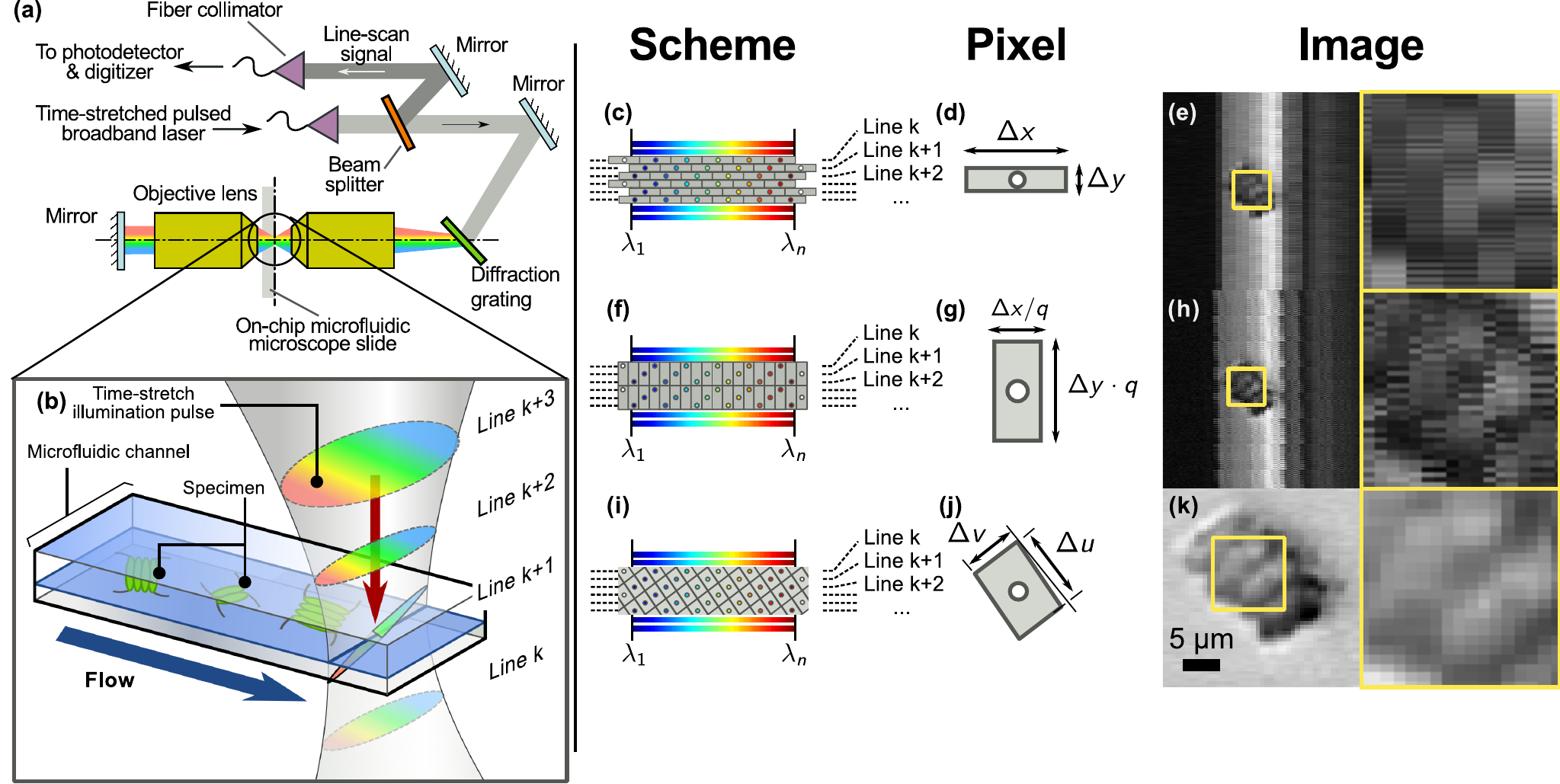}
\caption{%
{\bf Imaging flow cytometry setup with optical time-stretch capability.}\\
(a)~Imaging flow cytometry setup with optical time-stretch capability;
(b)~illustration of fast-axis scanning by spectral-encoding illumination and slow-axis scanning by ultrafast microfluidic flow;
(c--e)~conventional image restoration by aligning the time-stretch line-scans,
but disregarding the actual proximity of sampled points in neighboring line scans.
(f--h)~Interleaving multiple line-scans can resolve the high bandwidth time-stretch temporal waveform along the fast axis.
Both methods give rise to highly elongated pixels with aliasing along
the fast axis and the slow axis respectively.
(i--k)~Two-dimensional re-sampling utilizes relative subpixel drift $\delta x$ of neighboring line scans to interpolate from the same data. 
Even though the pixel area in panel~(g) is the same as that in panel~(d),
the spatial resolution improves along the fast axis after interpolation.
Insets: Zoom-in views of the restored optical time-stretch image.
}
\label{fig:schematic-principle}
\end{figure*}

We consider the most common form of time-stretch imaging that has been proven in a broad range of applications,
from flow cytometry to surface inspection,
i.e.\ on-the-fly line-scan imaging of the specimen [Fig.~\ref{fig:schematic-principle}(a--b)].
In this scenario, the pulsed and one-dimensional (1D) spectral shower illumination performs spectrally-encoded line-scanning of the unidirectional motion of the specimen,
e.g.\ biological cells in microfluidic flow (see \nameref{sec:methods}).
The two-dimensional (2D) image is reconstructed by digitally stacking the spectrally-encoded and time-stretched waveforms,
so that the fast axis of the resultant 2D image is the spectral-encoding direction,
and the slow axis corresponds to the flow direction [Fig.~\ref{fig:schematic-principle}(c)].

The pixel resolution along the flow direction (fast axis) is the product of linear flow speed $v_y$ and the laser pulse repetition rate $F$, i.e.\ $\Delta y = v_y/F$.
On the other hand, the pixel resolution along the spectral-encoding direction (slow axis) is independently determined by the resolving power of the imaging setup and that of the wavelength-to-time conversion,
i.e.\ $\Delta x = C_x(C_t f)^{-1}$,
where $C_x$ is the wavelength-to-space conversion factor of the spectral encoding setup;
$C_t$ is the wavelength-to-time conversion factor of the time-stretch spectrum analyzer;
and $f$ is the sampling rate of the digitizer.
When operated at a low sampling rate, time-stretch imaging of ultrafast flow generates highly elongated pixels [Fig.~\ref{fig:schematic-principle}(d)] that easily result in image aliasing [Fig.~\ref{fig:schematic-principle}(e)].
We find that the aspect ratio of the original LR image pixel, defined as
\begin{equation}
    r = \frac{\Delta y}{\Delta x}
    = \frac{v_y C_t}{C_x} \times \frac{f}{F},
    \label{eq:aspect-ratio}
\end{equation}
is as small as in the order of $10^{-2}$ in typical time-stretch imaging configuration (see the parameters described in \nameref{sec:methods}).
Ideally, if the sampling clock frequency $f$ of the digitizer is locked to the laser pulse repetition rate $F$, the line scans will align along the slow axis.
In practice, the average number of pixels per line scan ($=f/F$) is not an integer.
The line scan appears to ``drift'' along the slow axis,
and hence the image appears to be highly warped especially at low sampling rate [Supplementary Fig.~\ref{fig:algorithm}(b)].
Specifically, as the sampling rate $f$ is unlocked from the laser pulse repetition rate $F$,
pixel drift between adjacent time-stretch line-scans is observed,
and can be expressed as
\begin{equation}
    \delta x = \frac{C_x}{C_t} \times \left( \frac{1}{F} - N \times \frac{1}{f} \right),
    \label{eq:pixel-drifting}
\end{equation}
where integer $N$ is the number of pixels per line scan rounded off to the nearest integer.
It can be shown that $|\delta x| \leq \Delta x / 2$.
The warp angle is thus given as $\tan\theta = \delta x / \Delta y$, as illustrated in Supplementary Fig.~\ref{fig:algorithm}(b).
A common and straightforward approach to dewarp the image is to realign the digitally up-sampled line-scans.
However, this would, as shown later, result in image aliasing and artefact that are particularly severe at the lower sampling rate.
Furthermore, digital up-sampling of individual line scans does not provide additional image information and thus does not improve resolution along the fast axis.
An alternative approach is to interleave multiple line scans to resolve the high bandwidth 1D temporal waveform [Fig.~\ref{fig:schematic-principle}(f--h)].
It is commonly known as equivalent time sampling~\cite{KeysightTechnologies2015,Tektronix2001}.
However, fusion of multiple line-scans comes with the reduction of pixel resolution along the slow axis, which also introduces image aliasing.

We propose a pixel-SR strategy to harness this warping effect for creating the relative ``subpixel shift'' on both the fast and slow axes,
and thus restoring a high-resolution 2D time-stretch image [Fig.~\ref{fig:schematic-principle}(i--k)].
We first register the exact warp angle $\theta$ of the 2D grid [Supplementary Fig.~\ref{fig:algorithm}(e)].
It takes advantage of the non-uniform illumination background of the line-scan (mapped from the laser spectrum) as the reference,
thanks to the superior shot-to-shot spectral stability offered by the broadband mode-locked laser~\cite{Wei2015,Wei2014,Lau2016a}.
The precision of the measured warp angle critically influences the performance of the pixel-SR algorithm~\cite{Lin2004,Park2003}.
Next, the illumination background is suppressed by subtracting the intermediate ``dewarped'' image [Supplementary Fig.~\ref{fig:algorithm}(c)] with the high-bandwidth 1D reference illumination signal,
which is in turn restored by interleaving the first~$q$ LR time-stretch line-scans [Supplementary Fig.~\ref{fig:algorithm}(d)].
The 1D interleaving operation is based on a fast shift-and-add algorithm~\cite{Elad2001} together with rational number approximation. 
Finally, the image is denoised and re-sampled into the regular high-resolution grid~\cite{Lam2003},
thus reveals high-resolution information [Supplementary Fig.~\ref{fig:algorithm}(e)].
Detailed steps of the complete pixel-SR algorithm are included in the \nameref{sec:supplement}.

Note that interpolation of neighboring line-scans effectively enlarges the pixel size along the slow axis and reduces the effective imaging line-scan rate.
As shown in Fig.~\ref{fig:schematic-principle}(g),
the dimensions of the interpolated pixel along the warped direction are given as
\begin{subequations}
\begin{align}
    \Delta u &= \Delta x \cos \theta \\
    \Delta v &= \Delta y ( \cos \theta )^{-1}.
\end{align}%
\label{eq:transformed-pixel}%
\end{subequations}
This transform apparently does not resolve problem of aliasing because of the invariant pixel area,
i.e.\ $\Delta u\,\Delta v = \Delta x\,\Delta y$ for all $|\theta|<\pi/2$.
Nevertheless, when we consider the ratio of pixel size reduction, given as
\begin{subequations}
\begin{align}
    2r & < \frac{\Delta u}{\Delta x} \leq \frac{1}{\sqrt{2}} \\
    \sqrt{2}\times r & \leq \frac{\Delta v}{\Delta x} < \frac{1}{2},
\end{align}
\end{subequations}
the resolution improvement in the demonstration is particularly significant for highly elongated pixels [Eq.~\eqref{eq:aspect-ratio} and Fig.~\ref{fig:schematic-principle}(d,g)],
and a large warping ($|\tan \theta| \geq 1$).
Both cases are achievable at high repetition rate of ultrafast pulsed laser source (i.e.\
$F \gg v_y C_t C_x^{-1} f\approx 10^5\,\si{\hertz}$),
without compromising the overall imaging speed or throughput.
Also, the enlarged pixel size along the slow axis after interleaving is still well beyond the optical diffraction limit.
Note that the restored image resolution, as opposed to pixel resolution defined in Eq.~\eqref{eq:transformed-pixel}, is ultimately limited by the optical diffraction limit and the analog bandwidth of the digitizer.
In practice, these resolution limits are utilized to construct a matched filter to suppress noise from multiple low-resolution images in our pixel-SR algorithm (see \nameref{sec:supplement}).

\section{Results}
\label{sec:results}

\subsection{Pixel-SR time-stretch imaging of phytoplankton}

\begin{figure*}[tb]
\centering\includegraphics[width=\textwidth]{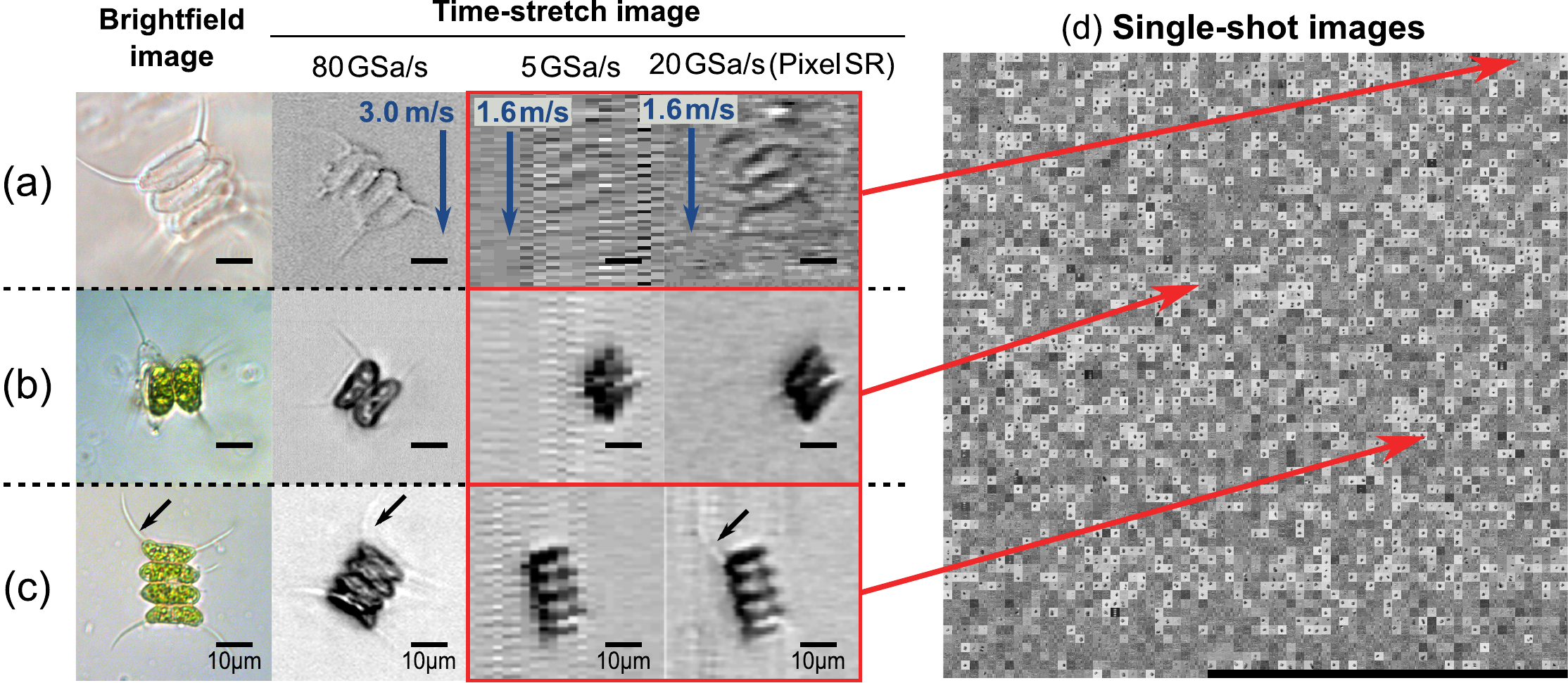}

\caption{%
{\bf \emph{Scenedesmus} samples captured by pixel-SR time-stretch imaging.}\\
Comparison of images in high resolution ($\SI{80}{\giga Sa\per\second}$,
$\SI{0.2}{\micro\meter\per pixel}$),
low resolution ($\SI{5}{\giga Sa\per\second}$,
$\SI{3.6}{\micro\meter\per pixel}$) and
pixel-SR (equivalent to $\SI{20}{\giga Sa\per\second}$,
$\SI{0.9}{\micro\meter\per pixel}$) for different cell sub-types:
(a)~discarded exoskeleton;
(b)~colonies with two daughter cells;
(c)~colonies with four daughter cells.
Refer to Supplementary Fig.~\ref{fig:more-comparison} for more examples.
(d)~Image collage of all 5,000 colonies and fragments acquired at $\SI{5}{\giga Sa\per\second}$.
Interactive version is available online at
\url{http://www.eee.hku.hk/~cschan/deepzoom/}.}
\label{fig:comparison}
\end{figure*}

To demonstrate pixel-SR for ultrafast time-stretch imaging with improved spatial resolution,
we chose a class of phytoplankton, \emph{scenedesmus} (Carolina Biological, USA), for its distinct morphological property.
\emph{Scenedesmus} is a colony of either two or four daughter cells surrounded by the cell wall of the mother (Fig.~\ref{fig:comparison}).
Each daughter cell possesses an elongated shape at around $\SI{5}{\micro\meter}$ in diameter along the minor-axis~\cite{Griffiths2010}.
Therefore, it serves as a model specimen to test resolution enhancement beyond
$\Delta x\approx \SI{2}{\micro\meter}$.

In the experiment, individual \emph{scenedesmus} were loaded into the microfluidic channel at an ultrafast linear flow velocity of $\SI{1}{\meter\per\second}$ to $\SI{3}{\meter\per\second}$,
which was manipulated based on the inertial flow focusing mechanism~\cite{DiCarlo2007,Martel2014}.
Time-stretch imaging of \emph{scenedesmus} was performed at a line-scan rate of $\SI{11.6}{\mega\hertz}$,
determined by the repetition rate of a home-built mode-locked laser (at center wavelength of $\SI{1.06}{\micro\meter}$).
The wavelength-time mapping was performed by a dispersive fiber module with a GVD of $\SI{0.4}{\nano\second\per\nano\meter}$,
which is sufficiently large to satisfy the ``far-field'' mapping condition,
i.e.\ the spatial resolution is not limited by GVD~\cite{Tsia2010}.
The time-stretch waveforms were then digitized by a real-time oscilloscope with adjustable sampling rate between $\SI{5}{\giga Sa\per\second}$ and $\SI{80}{\giga Sa\per\second}$.
Detailed  time-stretch imaging system configuration and experimental parameters are described in \nameref{sec:methods}.
At the highest possible sampling rate ($\SI{80}{\giga Sa\per\second}$), the cellular images comes with sharp outline and visible intracellular content (second column, Fig.~\ref{fig:comparison}).
At a lower sample rate of $\SI{5}{\giga Sa\per\second}$, however, the pixel dimensions become respectively $(\Delta x,\Delta y) = (\SI{3.6}{\micro\meter}, \SI{0.18}{\micro\meter})$.
As the diffraction limited resolution is estimated to be $\approx \SI{2}{\micro\meter}$,
the cell images captured at such a low sampling rate become highly aliased (third column, Fig.~\ref{fig:comparison}).

Together with the fact that the sampling clock was unlocked from the laser pulse frequency,
resolution enhancement by pixel-SR, i.e.\ to achieve a pixel size smaller than $\Delta x$, can thus be adopted in this scenario.
To support the above argument, we estimate the theoretical resolution improvement in this experimental setting.
From Eq.~\eqref{eq:pixel-drifting}, the relative pixel drifting is known to be roughly $\delta x = \SI{-1.8}{\micro\meter}$.
The warping of the 2D grid is given as $\tan\theta\approx -10$.
Therefore, the pixel-SR scheme can theoretically achieve $(\cos\theta)^{-1}\approx 10$~times improvement in resolution.
Note that in our current setup, the actual resolution improvement is roughly limited to 4~times,
i.e.\ at pixel size of $\SI{0.9}{\micro\meter}$ and at effective sampling rate of $\SI{20}{\giga Sa\per\second}$.
This is not inherent to the pixel-SR algorithm.
Instead, the restored image resolution is currently limited by the built-in signal conditioning filter at $\SI{10}{\giga\hertz}$ cut-off frequency in the oscilloscope,
implying that the additional subpixel measurements does not equally provide 10~times improvement in spatial resolution.

The warped grid is subsequently re-sampled to a regular rectangular grid at a pixel dimensions of $\SI{0.9}{\micro\meter} \times \SI{0.9}{\micro\meter}$.
The value of the warp angle $\theta$ is further refined by computational optimization (see \nameref{sec:supplement}) to ensure accurate pixel registration~\cite{Park2003}.
A spatial averaging filter is constructed to match the estimated optical diffraction limit and electronic filter bandwidth to average out the excess measurements and to suppress noise.
The restored pixel-SR images of the corresponding cell types are shown in Fig.~\ref{fig:comparison} and Supplementary Fig.~\ref{fig:more-comparison}.
The individual daughter cells in the \emph{scenedesmus} colonies are now clear of aliasing artifacts and noise.
Specifically, the hair protruding at the cell body, that is otherwise missing in LR time-stretch image, is now visible in the restored HR image.

\subsection{Morphological classification of phytoplankton}

Detailed spatial information of the cells
(i.e.\ size, shape and sub-cellular texture) can be exploited as the effective biomarkers for revealing cell types, cell states and their respective functions~\cite{Lau2016,Chen2016}.
Furthermore, such morphological information of cells can readily be visualized and analyzed by label-free optical imaging,
i.e.\ without the concern of cytotoxicity and photobleaching introduced by the fluorescence labelling, not to mention the costly labelling and laborious specimen preparation work.
To this end, taking advantage of HR image restoration, pixel-SR time-stretch imaging is particularly useful to enable label-free, high-throughput cellular classification and analysis based on the morphological features,
that is not possible with standard flow cytometry.
Here, we performed classification of sub-types of \emph{scenedesmus} ($n = 5,000$) imaged by our optofluidic pixel-SR time-stretch imaging system
(sampled at $\SI{5}{\giga Sa\per\second}$).
The images of individual colonies are reconstructed by pixel-SR algorithm.
Next, the images are computationally screened with a brightness threshold,
and then measured by a collection of label-free metrics.
The highly parallelized image processing and analysis procedures are performed on the high-performance computing cluster [see \nameref{sec:methods} and Supplementary Fig.~\ref{fig:cluster}].
Specifically, we aim at proving the capability of classification of two-daughter colonies and four-daughter colonies based on the label-free pixel-SR images.
Since the two variants belong to the same species, 
they serve as the relevant test subjects for label-free morphology-based cell classification.

\begin{figure*}[tb]
\includegraphics[width=\textwidth]{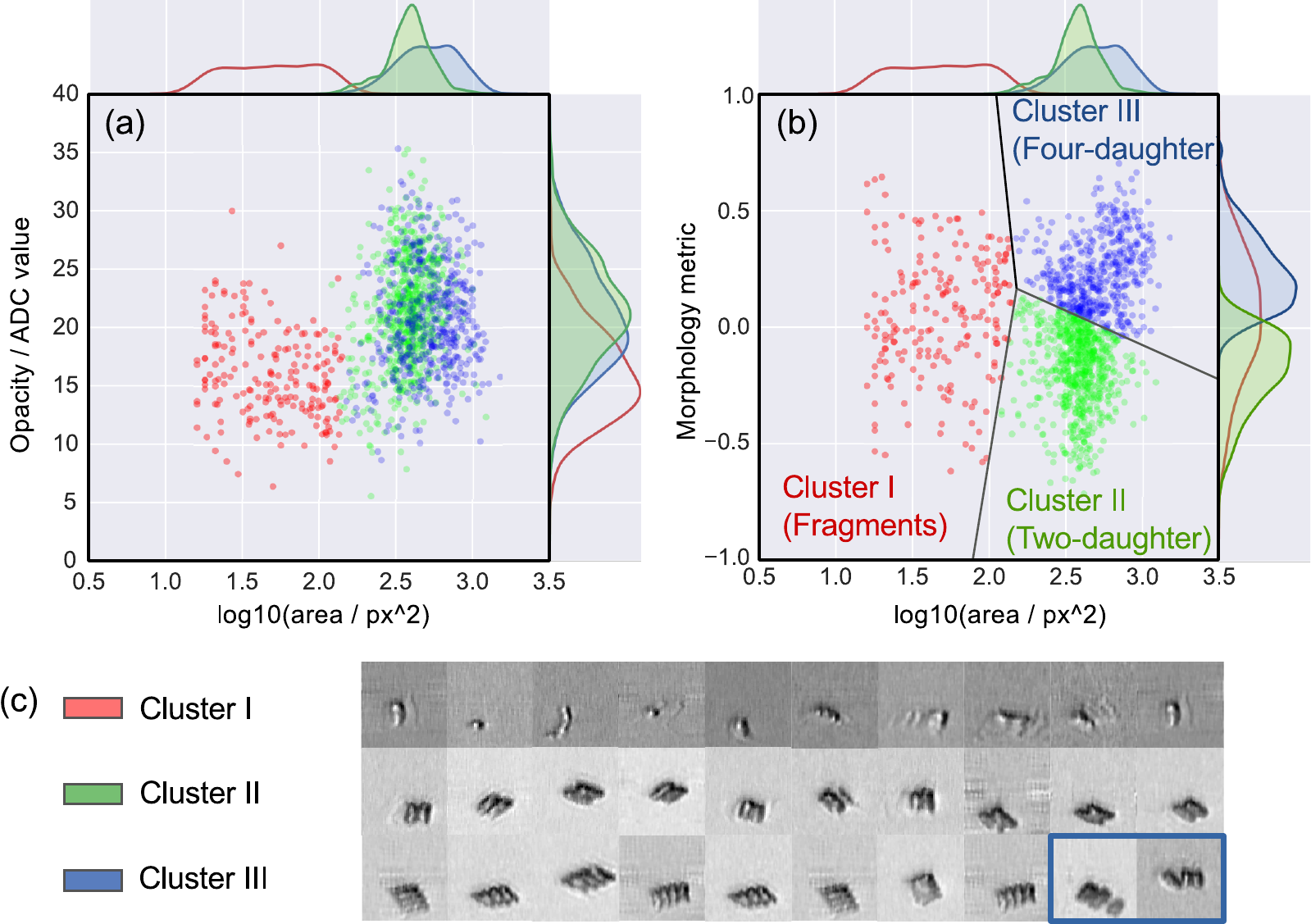}
\caption{
{\bf Classification of \emph{scenedesmus} samples based on
opacity, area, and morphology.}\\
(a)~Bulk metrics, i.e.\ opacity and area, are computed from time-stretch images restored by pixel-SR.
(b)~Classification is improved due to the morphology metric computed from high resolution image.
The histograms in~(a) and~(b) shows the distribution comparison of the various groups,
with and without pixel-SR.
(c)~HR image of the cell samples selected from the corresponding clusters.
(Bottom right, highlighted) The aggregates of smaller colonies are mis-classified as four-daughter colony, but is clearly distinguishable in pixel-SR time-stretch imaging.
The $1,368$ samples in the scatter plots were pre-screened from the $5,000$ pixel-SR image frames with the brightness threshold.
Interactive version is available online at
\url{http://www.eee.hku.hk/~cschan/scatter_plot/}.}
\label{fig:scatter-plot}
\end{figure*}

We first retrieved two label-free metrics of single cells: opacity and area from the restored pixel-SR frames
(see \nameref{sec:methods}).
These spatially-averaged metrics represent the optical density (attenuation) and the physical size of the \emph{scenedesmus} colonies respectively.
Based on the scatter plot of the screened samples ($n' = 1,368$) with the two metrics [Fig.~\ref{fig:scatter-plot}(a)],
fragments are easily distinguishable from the live cells because they are significantly smaller and more translucent [see also the images in Fig.~\ref{fig:scatter-plot}(c)].
Although it was conceived that the size of four-daughter colonies should be roughly double compared with that of the two-daughter colonies,
neither the area nor the opacity can be used to separate the two groups,
which appear to be highly overlapped in the scatter plot.
Clearly, these spatially-averaged metrics (or essentially LR metrics) failed to account for the subtle morphological differences between the two-daughter and four-daughter colonies,
both of which exhibit high variability in both the area and the opacity.

Next, the new morphology metric was extracted from the collections of pixel-SR images,
and plotted against cell area in the scatter plot [Fig.~\ref{fig:scatter-plot}(b)].
We encoded the morphological features of each image into the histogram of oriented gradients (HoG)~\cite{Dalal2005},
which was then projected to the most significant component using principle component analysis (PCA).
Essentially, this metric provides a measure of structural complexity of the cell bodies of the \emph{scenedesmus} colonies,
and thus produces better cluster separation compared to opacity metric,
with the four-daughter colonies distributed at larger morphology values [i.e.\ Cluster~III in Fig.~\ref{fig:scatter-plot}(b)] and two-daughter colonies at smaller values [i.e.\ Cluster~II in Fig.\ref{fig:scatter-plot}(b)].
The corresponding pixel-SR images are also randomly selected from each cluster for visual inspection,
the result of which indicates a good agreement between the classified results and the manually identified groups [Fig.~\ref{fig:scatter-plot}(c)].
We provide an interactive plot of Fig.~\ref{fig:scatter-plot}(b) that reveals the complete gallery of the HR images of each cluster at higher zoom level
(accessible online at \url{http://www.eee.hku.hk/~cschan/scatter_plot/}).

In practice, multiple morphological metrics can be measured from the pixel-SR image to improve classification accuracy~\cite{Chen2016}.
Owing to the discrete structure of \emph{senedesmus} colonies, it is possible to separate the two sub-types: two-daughter and four-daughter colonies more effectively in a higher-dimensional morphology metric space. 
For the sake of demonstration, only the first principle component of the morphological feature set, having the largest variance, is computed here as the morphology metric for classification.

\begin{figure*}[bph]
\centering\includegraphics[width=\textwidth]{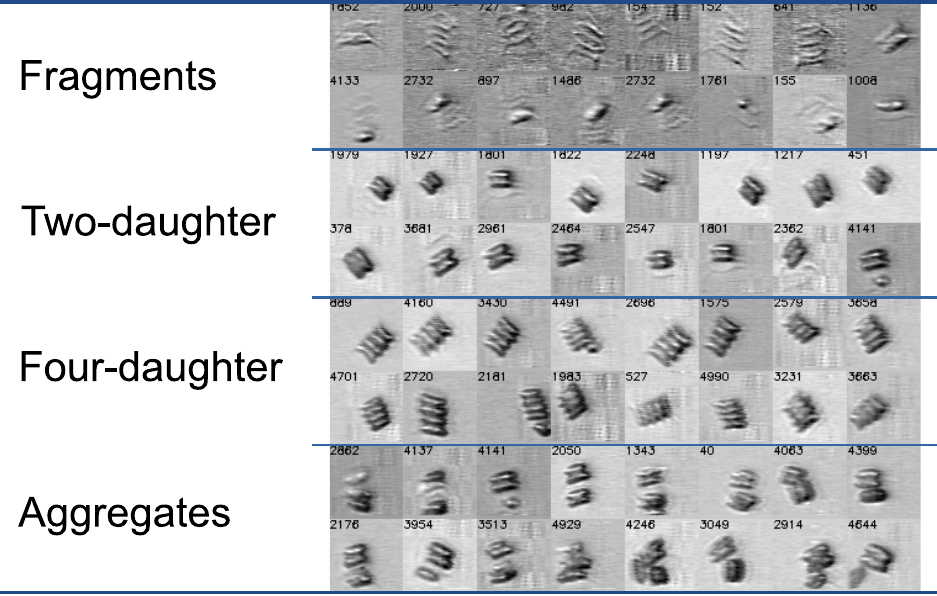}
\caption{%
{\bf Further examples of pixel-SR time-stretch image restoration of \emph{scenedesmus}.}\\
The four-digit number at the top-left corner of each image indicates the $i$-th image frame captured by the optofluidic time-stretch microscope setup.
Effective sampling rate: $\SI{20}{\giga Sa\per\second}$.
Image scale: $\SI{53}{\micro\meter} \times \SI{53}{\micro\meter}$.}
\label{fig:more-comparison}
\end{figure*}

We note that the present classification primarily focuses on the two populations,
i.e.\ two-daughter and four-daughter colonies.
Nevertheless, we observe that pixel-SR time-stretch images reveal further heterogeneity within the same population of phytoplankton.
Specifically, we identify a group of highly translucent fragments with well-defined exoskeleton structures (Supplementary Fig.~\ref{fig:more-comparison}),
and some rare aggregates as marked in blue frames [Fig.~\ref{fig:scatter-plot}(c) and Supplementary Fig.~\ref{fig:more-comparison}] ---
again demonstrating the imaging capability of pixel-SR time-stretch imaging for revealing rich morphological information at lower digitizer sampling rate.
Inadvertently, these translucent fragments and the aggregates are currently either computationally rejected as noise or mis-classified as four-daughter colonies,
introducing selection bias in the classification procedure.
In spite of this, the pixel-SR time-stretch images of all these outliers can be clearly identified by manual inspection alone.
More significantly, advanced automated non-linear object recognition algorithms,
such as artificial neural network for deep learning~\cite{Chen2016},
can be coupled with the present pixel-SR technique to improve the classification precision and sensitivity.

\subsection{Real-time continuous pixel-SR time-stretch imaging of microdroplets with field-programmable gate array (FPGA)}

As mentioned earlier, pixel-SR time-stretch imaging offers a practical advantage over direct acquisition at extremely high sampling rate, i.e.\ at $\SI{40}{\giga Sa\per\second}$ or beyond.
Ultrafast analog-to-digital conversion demands costly adoption of the state-of-the-art oscilloscopes that are conventionally equipped with limited memory buffers.
Not only does it hinder continuous, real-time on-the-fly data storage,
but also high-throughput post-processing and analytics.
Pixel-SR time-stretch imaging offers an effective approach to address this limitation by capturing the time-stretch images at a lower sampling rate (in the order of $\SI{1}{\giga Sa\per\second}$),
yet without compromising the image resolution.
More significantly, unlike the use of high-end oscilloscope in the previous experiments, a commercial-grade digitizer at a lower sampling rate can be readily equipped with an FPGA,
capable of continuous and reconfigurable streaming of enormous time-stretch image raw data to the distributed computer storage cluster (see \nameref{sec:methods}).
To demonstrate the applicability of pixel-SR to such a high-throughput data processing platform,
we performed continuous real-time monitoring of water-in-oil emulsion microdroplet generation in the microfluidic channel device at a linear flow velocity as high as $\SI{0.3}{\meter\per\second}$ and at the generation throughput of $5,800\,\mathrm{Droplet/s}$ (see \nameref{sec:methods}).
The time-stretch image signal is continuously recorded at the sampling rate of $f=\SI{3.2}{\giga Sa\per\second}$.

\begin{figure*}[tb]
\centering
\includegraphics[width=\textwidth]{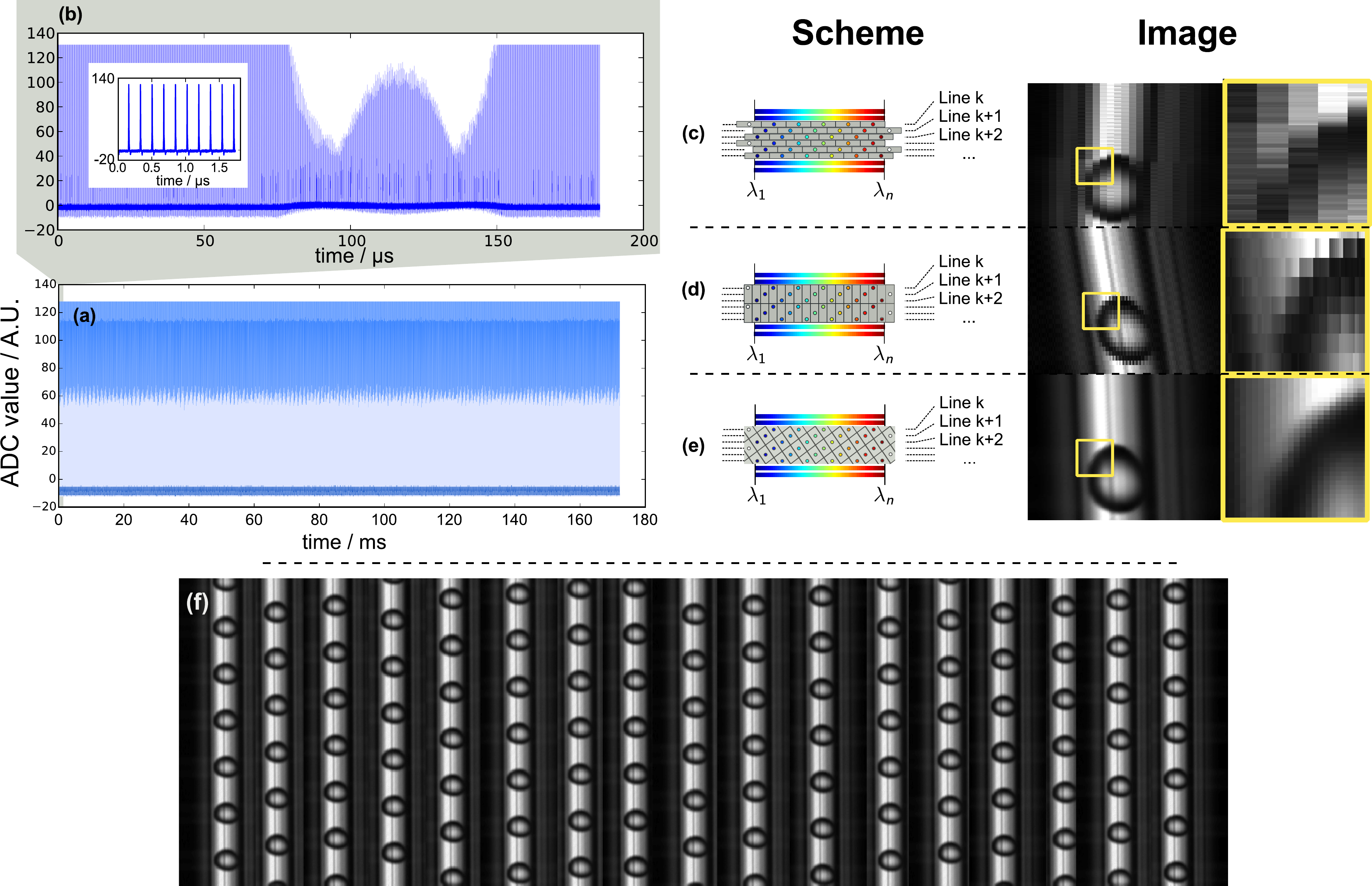}\\

\caption{%
{\bf Continuous acquisition of water-in-oil microdroplet at $\SI{3.2}{\giga Sa\per\second}$.}\\
Serialized time-stretch signal of (a)~$\SI{170}{\milli\second}$ duration, and
(b)~the first $\SI{200}{\micro\second}$.
The inset shows the first 10 line scans of the measurement.
Image restoration by
(c)~Pulse-by-pulse alignment.
(d)~One-dimensional equivalent time sampling.
(e)~Pixel super-resolution result.
The zoomed-in view of the images are also shown on the far right.
(f)~Snapshots of continuous recording of water-emulsion droplet imaging at a regular interval of $\SI{10}{\milli\second}$.
Totally 978 droplets are captured within the $\SI{170}{\milli\second}$ duration.}
\label{fig:water-droplet}
\end{figure*}

Similar to the previous observations with the oscilloscope, the raw time-stretch image captured by the commercial-grade digitizer is highly warped because of the timing mismatch between the laser and the digitizer [Fig.~\ref{fig:water-droplet}(b--c)].
It is reminded that the frequently adopted approach is to dewarp the image by aligning the individual line scans.
As shown in Fig.~\ref{fig:water-droplet}(c), each line scan is digitally re-sampled to realign the pixels along the slow axis.
While this strategy used to work for oversampled time-stretch signal at $\SI{16}{\giga\hertz}$ bandwidth~\cite{Wong2014,Chung2016},
it does not perform well at the low sampling rate because of signal aliasing [Fig.~\ref{fig:water-droplet}(c)].
Again, interleaving multiple line-scans comes with the degradation of pixel resolution along the slow axis as shown in Fig.~\ref{fig:water-droplet}(d).
In contrast, our pixel SR algorithm is able to restore HR time-stretch image, avoiding aliasing on either fast and slow axes.
The resolution improvement is five times the apparent pixel size,
i.e.\ at effective sampling rate of $\SI{16}{\giga Sa\per\second}$
and at pixel resolution of $\SI{1.1}{\micro\meter}$
[Fig.~\ref{fig:water-droplet}(e--f)].

It is noted that all image restoration procedures are currently done offline.
Although only the first $\SI{170}{\milli\second}$ is captured in this experiment for the sake of demonstration,
the maximum number of image pixels in the continuous image capture can be scaled up to the total data capacity of the computer cluster.
In our system, each computer node is equipped with the $\SI{256}{\giga B}$ hard disk drive [Supplementary Fig.~\ref{fig:cluster}(a)].
In principle, our method can enable real-time morphological image capture and object recognition at giga-pixel capacity.

\section{Discussions}

The spatial resolution of ultrafast time-stretch imaging is closely tied with the temporal resolution during data capture,
especially the sampling rate of the digitizer.
This feature of space-time mapping implies an overwhelming requirement on ultrahigh sampling rate, and thus the state-of-art digitizer in order to avoid image aliasing.
Not only does such high-end digitizer incur prohibitive high cost, but it also lacks sufficient memory depth for high-throughput continuous data storage, processing and analytics.
These key challenges have been impeding the widespread utility of time-stretch imaging in high-throughput applications, ranging from machine-vision and quality control in industrial manufacturing to single-cell analysis in basic biomedical and clinical diagnosis.
It is worth mentioning that compressive sampling can be an alternative solution to combat image aliasing at lower sampling rate,
which utilizes pseudo-random illumination patterns to preserve HR information of the image~\cite{Bosworth2013,Bosworth2015}.
The same effect is also achieved by modulating the spectrally-encoded pulse after illumination~\cite{Chen2014,Guo2015}.
Although around two-order-of-magnitude reduction in digitizer bandwidth has been achieved with this technique~\cite{Bosworth2015,Guo2015},
it requires time-consuming iterative process to restore the HR image.

In contrast, we proposed and demonstrated a pixel-SR technique that could enhance the pixel resolution (i.e.\ anti-aliasing) of ultrafast time-stretch imaging at the lower sampling rate, which is largely supported by the commercial-grade digitizers.
Resembling the concept of equivalent time sampling that is employed high-speed sampling oscilloscopes,
our pixel-SR technique harnesses the fact that subpixel shift of consecutive time-stretch line scans is innately generated by the mismatch between laser pulse repetition frequency and sampling frequency ---
a feature appeared virtually in all types of time-stretch imaging modalities.
Therefore, it requires no active synchronized control of illumination or detection for precise sub-pixel shift operation at an ultrafast rate.

Incidentally, the sub-pixel registration --- a key to ensure the robustness of the pixel-SR reconstruction, has been challenging as the sub-pixel shift is very often arbitrary and is further complicated by the motion blur.
Thanks to the superior shot-to-shot line-scanning offered by the stable mode-locked laser as well as the motion-blur-free imaging guaranteed by the ultrafast line-scan (i.e.\ at MHz and beyond),
our technique allows high-precision sub-pixel shift estimation.
We have demonstrated the strength of this method by enhancing pixel resolution of existing time-stretch imaging flow cytometry setup without additional hardware.
In our experiments, we showed that the digitizer sampling rate can be relaxed to $\SI{5}{\giga Sa\per\second}$ with our technique from the original $\SI{80}{\giga Sa\per\second}$.
Cellular texture, which is otherwise obscured in the LR time-stretch images,
can be restored with the pixel-SR algorithm.
More importantly, the restored HR time-stretch images enables better classification of biological cell sub-types.
Notably, we have also implemented the pixel-SR time-stretch imaging technique with the high-throughput data-acquisition platform based on FPGA at the sampling rate of $\SI{3.2}{\giga Sa\per\second}$.
Note that the compatibility of the pixel-SR algorithm to the FPGA is significant,
in that the integration of both can represent a cost-effective and practical solution for a wide variety of high-throughput time-stretch imaging applications.
While this pixel-SR technique is currently demonstrated in the context of time-stretch imaging, this concept can be generally applicable to any ultrafast line-scan imaging modalities with a single-pixel detector,
where the asynchronous sampling is involved in the image data capture.

\begin{figure*}[p]
\centering\includegraphics[width=\textwidth]{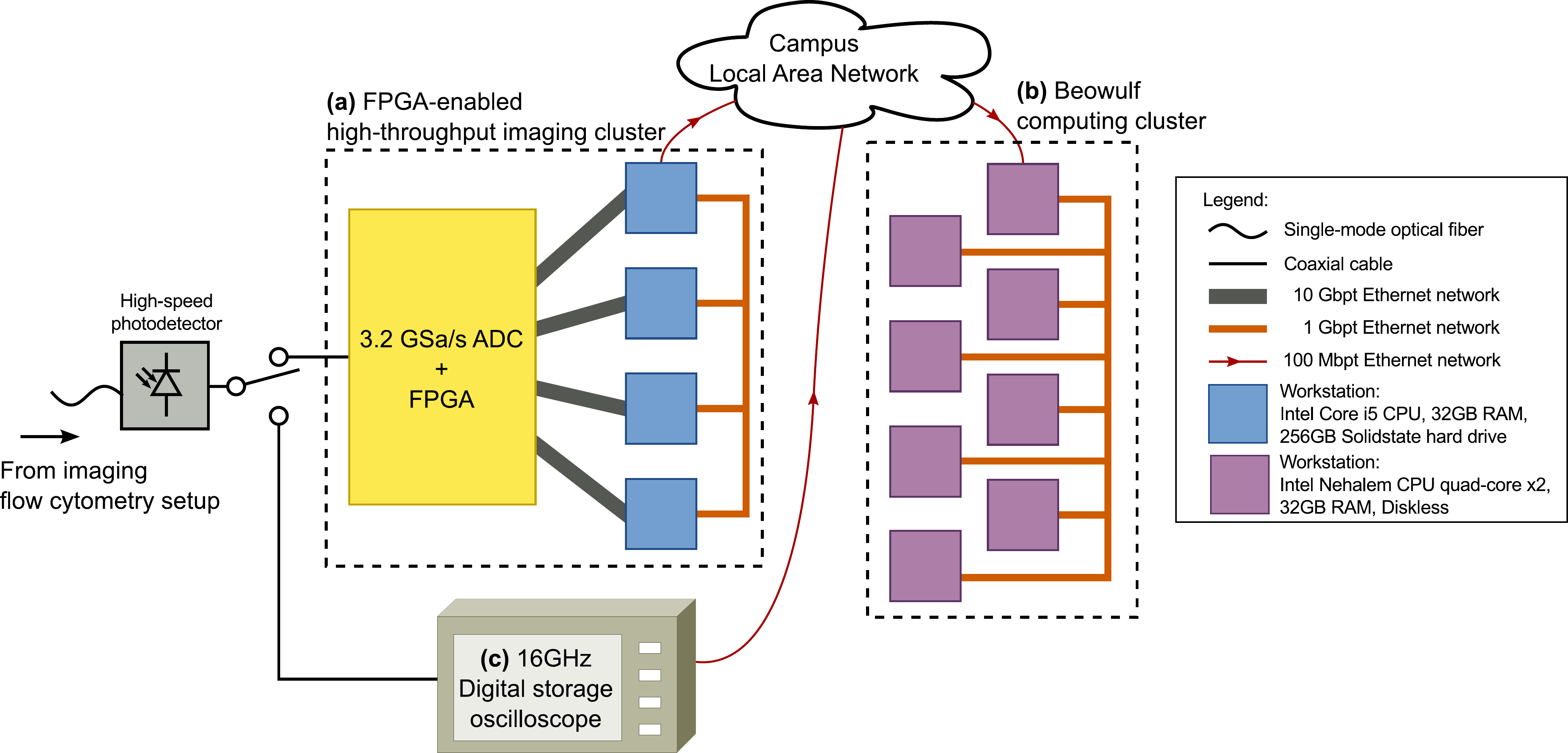}

\caption{%
{\bf Computing architecture for digital acquisition and analysis of time-stretch image signal.}\\
(a)~digitizer with field-programmable gate array capable of giga-pixel time-stretch imaging;
(b)~high-performance computing cluster for parallel time-stretch image restoration and analysis;
(c)~high-end oscilloscope for image resolution comparison.}
\label{fig:cluster}
\end{figure*}

In this paper, the interpolation algorithm is executed on a single processing core for each 1D time-stretch data segment representing one image frame;
the 5,000 image frames of \emph{scenedesmus} colonies (see \nameref{sec:methods}) are restored independently in multiple cores of the high-performance computing cluster [Supplementary Fig.~\ref{fig:cluster}(b)]
to achieve a real-time combined data crunching rate of $\SI{26.0}{\mega Sa\per\second}\approx 104\text{~frames per second}$.
Conceptually, the pixel-SR algorithm can be implemented in the graphical processing unit (GPU)~\cite{Beets2000,nVIDIA2001} as a massively parallel routine to increase the data crunching rate by up to two orders-of-magnitude,
i.e.\ in the order of $\SI{10}{\giga Sa\per\second}$.
As mentioned before, this algorithm can be programmed in the FPGA for real-time image restoration and classification,
further eliminating the back-end computation resources depicted in Supplementary Fig.~\ref{fig:cluster}(b).
This will be the next stage of our recent work on real-time \emph{in situ} classification~\cite{Chung2016}.

\section{Methods}
\label{sec:methods}

\subsection{Optofluidic time-stretch microscopy system}

Figure~\ref{fig:schematic-principle}(a) shows the schematics of the optofluidic time-stretch microscope with a double-pass configuration~\cite{Sheppard1980}.
The microfluidic channel is illuminated by a spectrally-encoded pulsed laser beam (center wavelength$=\SI{1060}{\nano\meter}$; bandwidth$=\SI{20}{\nano\meter}$).
As the biological cells or microparticles travel along the microfluidic channel, a train of time-stretched illumination pulses captures a sequence of line-scans across the cell at a laser pulse repetition rate of $F=(11.6142 \pm 0.0005)\si{\mega\hertz}$.
The detection module, which consists of a $\SI{12}{\giga\hertz}$ photodetector (1544-B, Newport) and the $\SI{33}{\giga\hertz}$-bandwidth digital storage oscilloscope (DSAV334A, Keysight Technologies),
then records and digitizes the captured line-scan sequence at the instaneous sampling rate from $\SI{5}{\giga Sa\per\second}$ to $\SI{80}{\giga Sa\per\second}$ [Fig.~\ref{fig:schematic-principle}(b)].
The infinity-corrected microscope objective lens (L-40X, Newport) at numerical aperture of $0.66$ and the transmission diffraction grating  (WP-1200, Wasatch Photonics) at $\SI{1200}{groove\per\milli\meter}$ achieve the wavelength-to-space conversion factor of
$C_x = \SI{7.1}{\micro\meter\per\nano\meter}$.
The wavelength-to-time conversion factor $C_t= \SI{400}{\pico\second\per\nano\meter}$ is achieved with a $\SI{5}{\kilo\meter}$ long single-mode dispersive fiber (1060-XP, Nufern).
The values of both conversion factors are kept fixed in our experiments.

\subsection{Imaging flow cytometry protocol}

A population of more than 10,000 units of phytoplankton is loaded into the microfluidic channel with a syringe pump (PHD22/2000, Harvard Apparatus) at a linear speed of $\SI{1.6}{\meter\per\second}$,
A polydimethylsiloxane (PDMS) microfluidic channel is designed and fabricated based on ultraviolet (UV) soft-lithography.
Detailed fabrication steps has been described in Ref.~\cite{Wong2014}.
The width ($\SI{60}{\micro\meter}$) and height ($\SI{30}{\micro\meter}$) of the channel are chosen such that the balance between the inertial lift force and the viscous drag force is achieved for manipulating the positions of the individual cells (with the size of $\approx5\text{--}\SI{20}{\micro\meter}$) and focusing them in ultrafast flow inside the channel~\cite{DiCarlo2007,Martel2014}.

The time-stretch signal is first captured and digitized by the oscilloscope (DSAV334A, Keysight Technologies) of maximum analog bandwidth of $\SI{33}{\giga\hertz}$.
Without changing hardware, the sampling rate of the oscilloscope is down-adjusted from
$\SI{80.000}{\giga Sa\per\second}$ to $\SI{5.000}{\giga Sa\per\second}$.
At lower sampling rate, it is known to possess a signal conditioning filter at $\SI{10}{\giga\hertz}$ cut-off frequency.
Because of the limited memory depth of the oscilloscope, the time-stretch waveform is captured in segmented mode,
where only 5,000 units are captured during the experiment [Fig.~\ref{fig:comparison}(a)].
More examples of pixel-SR images are shown in
Supplementary Fig.~\ref{fig:more-comparison}.

\subsection{Morphological classification methods}

For classification of unlabelled cell by morphology, we first attempt to group the cells in terms of cell mass and volume.
These metrics are represented by cell opacity and area respectively, both measured from each single-colony image captured from the experiment [Fig.~\ref{fig:scatter-plot}(a)].
\emph{Opacity} is computed by taking the average of all pixel values, before subtracting the background pixel value.
\emph{Area}, on the other hand, is computed from the rotated rectangular box enclosing the cell in the image.
The rectangular box is tightly fitted to the outline of the external exoskeletons of the \emph{scenedesmus},
which in turn is extracted by a brightness threshold filter,
to minimize the area.
The samples with zero area, i.e.\ cells/fragments that are nearly transparent, are screened out before the classification.
To compute the morphological metrics, the cell bodies of the \emph{scenedesmus} is first cropped, rotated and then scaled with the tightly-fitted rectangular box that is obtained earlier.
This step is to ensure that the morphological metric would be invariant to scale, rotation and aspect-ratio of the cell body.
A set of histogram of oriented gradients measurements (HoG) is then computed from the intermediate image,
which is then projected to the most significant component using principle component analysis (PCA).
As mentioned in the discussions,
the cell samples are automatically classified into three disjoint clusters by the K-means clustering algorithm.
To avoid human bias, initial cluster centroids are generated from random coordinates in the morphological metric versus log10(area) space, as presented in Fig.~\ref{fig:scatter-plot}(b).
The exact names of the three clusters (i.e.\ cell fragments, two-daughter colony and four-daughter colony)
are later identified by inspecting the pixel-SR images in each cluster.

\subsection{Continuous high-throughput imaging of emulsion generation with field-programmable gate array (FPGA)}

In Fig.~\ref{fig:water-droplet}, the water-in-oil emulsion microdroplets are generated \emph{in situ} in the PDMS-based microfluidic water injection device~\cite{Anna2003,Kim2010},
which is mounted on the imaging flow cytometry setup.
The water microdroplets are in laminar flow at a linear speed of $\SI{0.3}{\meter\per\second}$.
The time-stretch signal is digitized by the $\SI{3.2000}{\giga Sa\per\second}$ 8-bit analog-to-digital converter (ADC) (a custom design by the Academia Sinica Institute of Astronomy and Astrophysics, Taiwan),
and is subsequently distributed in real-time to four computing workstations by a field-programmable gate array (FPGA) (Virtex-6 SX475T, Xilinx\cite{CASPER2016})
The digital acquisition system is illustrated in Fig.~\ref{fig:cluster}(a).
Since each workstation is equipped with the $\SI{256}{\giga B}$ solid-state hard drive, the system is capable of continuous high-throughput recording in the order of $10^{3}\,\si{\giga Pixels}$.
For the sake of demonstration, only the first $\SI{160}{\milli\second}$ is recorded.

\section*{Author contribution}

Conceived and designed the algorithm/experiments:
A.C.S.C.
Contributed reagents/materials/analysis tools:
H.C.N.,
S.C.V.B.,
Wrote the paper:
A.C.S.C.,
E.Y.L.,
K.K.T..
Supervised the project:
H.K.H.S.,
E.Y.L.,
K.K.T..

\section*{Acknowledgments}

We thank Queenie T.~K.\ Lai for providing the phytoplankton culture, and
Bob~C.~M.~Chung for fabricating the on-chip water-injection and microfluidic inertial flow focusing microscope slides.
We also thank Xing Xun for the technical support on unsupervised pattern recognition and classification algorithms.
This work is conducted in part using the HKU ITS research computing facilities that are supported in part by the Hong Kong UGC Special Equipment Grant (SEG HKU09).
This work is partially supported by grants from the 
Research Grant Council of the Hong Kong Special Administration Region, China (Project No. 17208414, 717212E, 717911E, 17207715, 17207714, 720112E),
Innovation and Technology Support Programme (ITS/090/14),
University Development Fund of HKU, and the
National Natural Science Foundation of China (NSFC)/Research Grants Council (RGC) Joint Research Scheme (N\_HKU714/13).

\clearpage
\myappendix
\section{Supporting information}
\label{sec:supplement}

\linenumbers
\subsection{General framework of pixel super-resolution algorithm}
\label{sec:algorithm}

The optofluidic time-stretch imaging process is modelled as
\begin{subequations}
\begin{align}
    g(x,y) &= h_1(x,y) \ast f(x,y) + I_B(x) \\
      I(t) &= h_2(t) \ast S[ g(x+y \tan\theta,y) ] + n(t),
\end{align}%
\label{eq:forward-model}%
\end{subequations}
where $I(t)$ is the distorted low-resolution measurement of the object $f(x,y)$ at the presence of measurement noise $n(t)$;
and $g(x,y)$ is the intermediate image of the object illuminated by the spectrally-encoded line beam $I_B(x)$.
Functions $h_1(x,y)$ and $h_2(t)$ correspond to the 2D point spread function (PSF) of the optical system and the 1D signal pre-conditioning filter of the digitizer respectively.
The warp angle $\theta$ accounts for the image distortion brought by the asynchronous sampling of the time-stretch pulse train.
The serialization operator $S(\cdot)$, representing the line-scan process, uniquely maps each spatial coordinates $(x,y)$ to time $t$.
The continuous 1D signal $I(t)$ is subsequently sampled by the digitizer.
Here, we seek to restore the object $f(x,y)$ from the time-stretch measurement $I(t)$ with the pixel-SR algorithm.

\paragraph{Step 1: Signal de-serialization.}

\begin{figure*}[tb]
\centering\includegraphics[width=\textwidth]{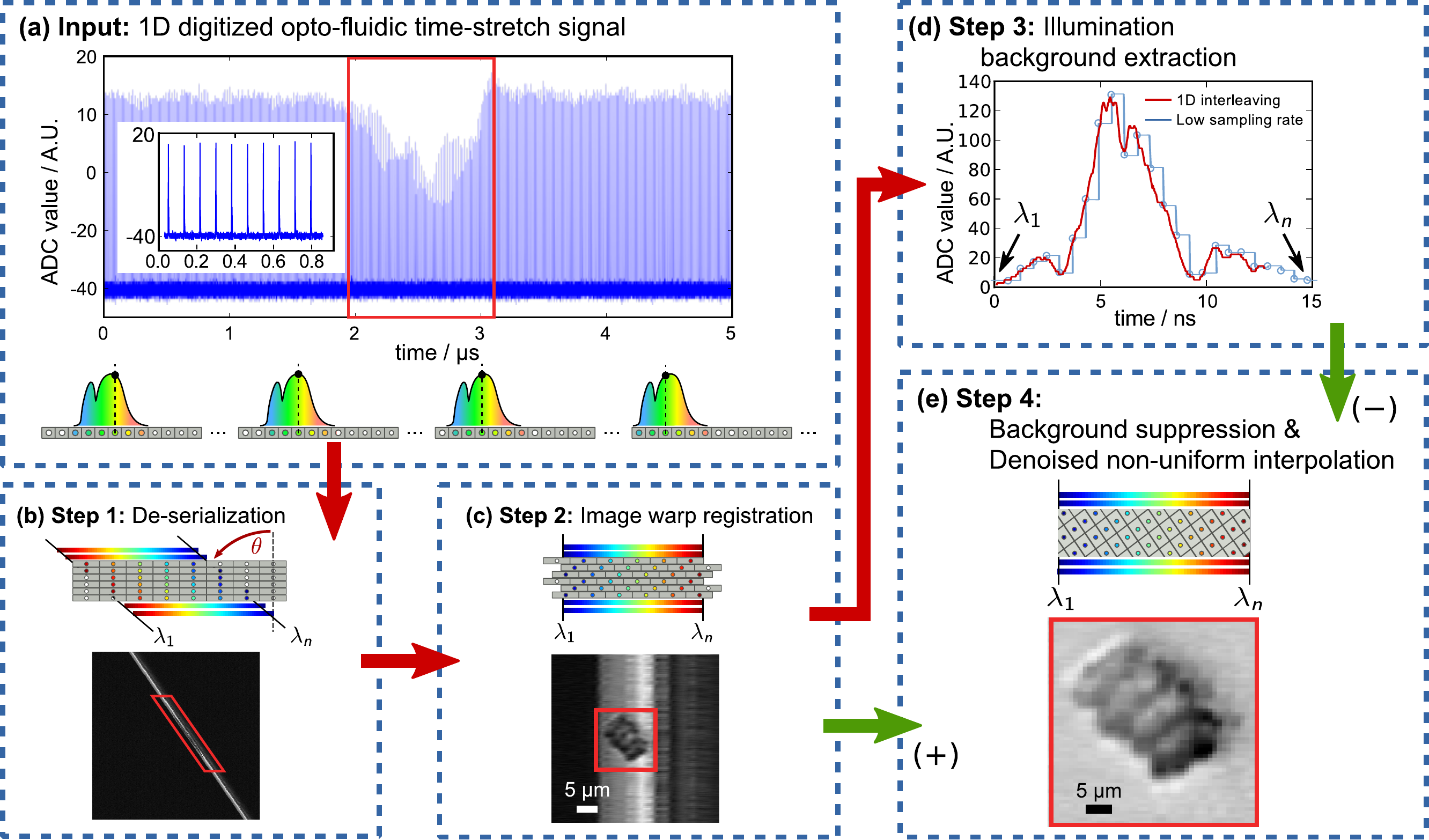}
\caption{%
{\bf Flow chart of the pixel-SR algorithm for optofluidic time-stretch microscopy.}\\
(a)~Serialized time-stretch signal of water emulsion droplet.
The inset shows the first 10~line scans of the measurement.
(b)~De-serializing the time stretch signal, showing the high warp angle $\theta$ due to pixel drift.
(c)~With alignment of the illumination pulse, the warp angle $\theta$ can be precisely estimated, so does the exact pixel coordinates in space.
(d)~Illumination background extraction by 1D equivalent time sampling of the first $q$~line scans.
(e)~Denoised non-uniform interpolation of the aligned measurement.}
\label{fig:algorithm}
\end{figure*}

Prior to high-resolution image restoration, each input 1D signal $I(t)$ [Supplementary Fig.~\ref{fig:algorithm}(a)] is first de-serialized to form an intermediate image $I(x,y)$,
which doesn't modify the digital samples [Supplementary Fig.~\ref{fig:algorithm}(b)].
By re-arranging the terms in Eqs.~\eqref{eq:forward-model}, the image corruption model can be represented as
\begin{equation}
    I(x,y) = W_\theta [ h(x,y) \ast f(x,y) + I'_B(x) ] + n(x,y),
    \label{eq:forward-model2}
\end{equation}
where $h(x,y)=h_1(x,y) \ast h_1(t=C_t C_x^{-1} x)$; and $I'_B(x) = h_2(x) \ast I_B(x)$.
The image warp transform operator $W_\theta[\cdot]$ maps the spatial coordinates such that $(x,y) \mapsto (x+y \tan \theta,y)$.
In the following paragraphs, we describe the methods to estimate the image distortion parameters from the captured data
[i.e.\ $W_\theta(\cdot)$, and $I'_B(x)$],
and subsequently restore the object $f(x,y)$ by denoised non-uniform interpolation.

\paragraph{Step 2: Pixel registration by image warp estimation.}

The performance of pixel-SR is highly sensitive to errors in pixel registration.
The initial value of the relative pixel drift $\delta x$ can be obtained from the specification of the pulsed laser and the digitizer.
However, its precise value can only be estimated from the captured data because the laser cavity length varies in accordance with ambient temperature and mechanical perturbation.
In our approach, we achieve accurate pixel registration by optimizing background suppression.
Compared to the moving object $f(x,y)$ that contributes to varying spectral shape in the time-stretch line scans,
the laser spectral shape $I'_B(x)$ is highly stable from pulse to pulse, and appears as straight bands in the background of the captured raw data.
Owing to the presence of pixel drift,
the line scans are warped at an angle $\theta$ [Supplementary Fig.~\ref{fig:algorithm}(b)].
This warping needs to be compensated for accurate extraction of the lasing spectrum, evaluated as
\begin{equation}
    I'_B(x)|_\theta \approx \frac{1}{M\Delta y} \int_0^{M\Delta y} W^{-1}_\theta [I(x,y)] \,dy,
    \label{eq:background}
\end{equation}
where $M$ is the number of line scans of the warped image $I$.
Ideally, a ``clean'' foreground [Supplementary Fig.~\ref{fig:algorithm}(f)] can be obtained by direct subtraction of the lasing background $I_B(x)$ from the dewarped image $W^{-1}_\theta[I(x,y)]$.
Error in the value of the warp angle $\theta$ induces distortions in the estimated illumination background $I_B(x)$, thus results in band-like artefacts superimposed onto the foreground object.
However, this property can be exploited to obtain an accurate value $\hat\theta$ by maximizing the ``cleaniness'' of the extracted foreground,
i.e. by minimizing the squared residual in the foreground, expressed as
\begin{equation}
    \hat\theta =\arg\min_{\theta} \int_0^{N\Delta x}\int_0^{M\Delta y} \left[
        W^{-1}_\theta [I(x,y)] - I'_B(x)|_\theta
    \right]^2\,dy\,dx,
\end{equation}
where the integer $N$ is the number of pixels of each line scan.
Supplementary Figure~\ref{fig:algorithm}(c)~depicts such image warp registration process.
The accuracy of this pixel registration approach is fundamentally limited by the ``decorrelation distance'' of the laser spectrum,
or alternatively named the spectral coherence of the time-stretched illumination pulse, i.e. $\epsilon[\tan\theta] < C_x \delta \lambda / (M\Delta y)$.

The estimated warp angle $\theta$ is then utilized to compute the spatial coordinates of all pixels in the low-resolution line scans $I(x,y)$, as shown in Supplementary Fig.~\ref{fig:algorithm}(c).
The registered pixel coordinates and the corresponding pixel value $(x_j, y_j, I_j)$ are then indexed in the $k$-dimensional (K-D)~tree structure~\cite{Maneewongvatana2001} for efficient searching.

\paragraph{Step 3: Illumination background extraction.}

The undulated laser spectrum extracted in Eq.~\eqref{eq:background}, i.e.\ $I'_B (x)|_{\theta=\hat\theta}$, is also aliased;
it must be restored to further suppress the illumination background.
This problem can be solved by interleaving the first several line scans, in which the object is absent [Supplementary Fig.~\ref{fig:algorithm}(d)].
A fast shift-and-add algorithm~\cite{Elad2001} is implemented to interleave the first $q$ low-resolution time-stretch pulses into the high-resolution 1D grid.
The time-stretch pulses are zero-filled and shifted before adding up the signals.
To enable fast pixelized operations,
the relative shift of the $k$-th pulse ($k\leq q$) is rounded to the multiple of $\Delta x/q$.
It is desirable for $q$ to be large to achieve a higher pixel registration.
However, time-interleaving of multiple pulses reduces the effective imaging line-scan rate.
In other words, the effective pixel size along the slow axis ($q\Delta y$) must be smaller than the optical diffraction limit to avoid image aliasing.
The optimization criteria of $q$ is formulated as
\begin{equation}
    \min_{0<q < r / \Delta y}
    \left\vert \frac{f}{F} - \left( N + \frac{p}{q} \right) \right\vert,
\end{equation}
where $p,q$ are integers;
integer $N$ is the number of pixels per line scan rounded off to the nearest integer;
$r$ is the diffraction limit of the optical microscopy system.
Mathematically, this problem is equivalent to the rational number approximation,
where the solution is the truncated continued fraction of $f/F$ computed from the Euclidean algorithm~\cite{Stillwell2003}.
The relative subpixel shift of the $k$-th pulse ($k\leq q$) is thus determined as $d_k = [p(k-1) \mod q]/q\times \Delta x$.

This time-interleaving algorithm is also applied in Fig.~\ref{fig:water-droplet}(d).
At the sampling rate of $\SI{3.2}{\giga Sa\per\second}$, the best approximation of $f/F$ is $(275\frac{16}{31})$,
i.e.\ $N=276$; $p=-15$; and $q=31$.
The low-resolution signals of the time-stretch pulses are thus shifted and added in the following locations:
\begin{equation*}
\{d_k\} = \left\lbrace 0,    \frac{16}{31}\Delta x,  \frac{1}{31}\Delta x,
\frac{17}{31}\Delta x,    \frac{2}{31}\Delta x,  \ldots,
\frac{15}{31}\Delta x
\right\rbrace.
\end{equation*}
The anti-aliased laser spectrum is given as~\cite{Ben-Ezra2005}
\begin{equation}
	\hat I'_B(x) = \sum_{k=1}^q I[x, (k-1)\Delta y]\times
	\mathrm{comb} \left( \frac{x-d_k}{\Delta x} \right),
\end{equation}
where $\mathrm{comb}(\cdot)$ is a train of impulse functions~\cite{Bracewell1986}.

\paragraph{Step 4: Denoised non-uniform interpolation.}

\def\vecf{\mathbf{f}}
\def\vecIp{\mathrm{I}'}
\def\vech{\mathrm{h}}
\def\matH{\mathbf{H}}
\def\matW{\mathbf{W}}

Pixel-SR restoration of the object $f(x,y)$ can be obtained in two stages: non-uniform interpolation and image denoising~\cite{Elad1997,Park2003}.
For higher computational efficiency, the above two stages are performed at once by utilizing the value of the denoising filter as the weights in the interpolation process [Supplementary Fig.~\ref{fig:algorithm}(e)].
Our objective is to restore $f(x,y)$ from Eq.~\ref{eq:forward-model2} by minimizing the noise $n(x,y)$, i.e.
\begin{align}
        \hat f(x,y) &=  \arg\min_f \int_0^{M\Delta y} \int_0^{N\Delta x} [ n(x,y) ]^2 \,dx\,dy \nonumber \\
                    &=  \arg\min_f \int_0^{M\Delta y} \int_0^{N\Delta x} 
    [ I(x,y) - \nonumber\\
    &\quad W_\theta\lbrace h(x,y) \ast f(x,y) + \hat I'_B(x)\rbrace  ]^2 \,dx\,dy
    \label{eq:SR-lsq}
\end{align}
For a digital signal $\vecIp$ and the discrete image
$\vecf=\{f_i: f_i=f(x_i,y_i)\}$, Eq.~\eqref{eq:SR-lsq} can be rewritten as
\begin{equation}
      \mathrm{\hat f} = \arg\min_\vecf \left\Vert \vecIp - \matH \vecf \right\Vert^2_2,
      \label{eq:SR-lsq2}
\end{equation}
where $\mathbf{W}_\theta$ and $\mathbf{H}$ are the matrix representation of the operator $W_\theta[\cdot]$ and the convolution kernel $h(x,y)$ respectively; and
$\mathrm{I}' = \mathbf{W}_\theta^{-1} \mathrm{I} - \mathrm{\hat I}'_B$ is the dewarped, background-suppressed time-stretch signal from Step~3.
Solving Eq.~\eqref{eq:SR-lsq2} directly is not feasible for ultrafast imaging application,
not only because of the sheer size of matrix $\mathbf{H}$,
but also of the potential noise amplification effect of the ill-conditioned problem~\cite{Lam2003}.
In practice, Eq.~\eqref{eq:SR-lsq} can be computed more efficiently by exploiting that fact that the kernel $h(x,y)$ is sparse.
That is, the target pixel area of the high-resolution image $f(x,y)$ is set to be just slightly smaller than the area of the 2D convolution kernel $h(x,y)$,
such that the kernel $h(x,y)$ at the location of $i$-the pixel $(x_i, y_i)$,
which corresponds to the $i$-th column of matrix $\matH$,
is only weakly correlated to that of the neighbouring pixels,
i.e. $| \vech_i^T \vech_j| \ll |\vech_i^T \vech_i|$ for all $j\neq i$.
This pixel size selection also achieves critical sampling of the high-resolution image $f(x,y)$.
Hence, the minimizer in Eq.~\eqref{eq:SR-lsq2} can now be approximated as
\begin{align}
   \min_\vecf \left\Vert \vecIp - \matH \vecf \right\Vert_2^2 
   &= \min_\vecf \left\Vert \vecIp - \sum_{j=1}^L \vech_i f_i \right\Vert_2^2 \nonumber \\
   &\approx \min_\vecf \sum_{i=1}^L \Vert \vecIp - \vech_i f_i \Vert_2^2 
   		- (L-1) \Vert I' \Vert_2^2 \\
   &\approx \sum_{i=1}^L \min_{f_i} \Vert \vecIp - \vech_i f_i \Vert_2^2
	    - (L-1) \Vert I' \Vert_2^2,
\end{align}
where $L$ is the total number of high-resolution pixels of image $f(x,y)$;
and $\vech_i$ is the $i$-th column of matrix $\matH$.
For the $i$-th element of $\vecf$ at registered coordinates $(x_i,y_i)$, its pixel value is
\begin{align}
    \hat f_i &\approx \arg\min_{f_i} \Vert \vecIp - \vech_i f_i \Vert_2^2 \nonumber \\
             &= \frac{\vech_i^T \vecIp}{| \vech_i |} \nonumber \\
     \Rightarrow \hat f(x_i,y_i)
     &\approx \frac{\sum_{j=1}^{MN} h(x_j-x_i, y_j-y_i)\times [I_j - \hat I_B(x_j)]}
     				{\sum_{j=1}^{MN} [h(x_j-x_i, y_j-y_i)]^2}
\label{eq:interpolation} \\
             &= \frac{\sum_{r_m \leq \sigma} h(x_m-x_i, y_m-y_i)\times [I_m - \hat I_B(x_m)]}
             		{\sum_{r_m \leq \sigma} [h(x_m-x_i, y_m-y_i)]^2},
\label{eq:interpolation-neighbour}
\end{align}
where pixel value $I_m$ corresponds to the $m$-th nearest neighbour of the spatial coordinate $(x_i, y_i)$;
and distance $r_m = \sqrt{ (x_i - x_m)^2 + (y_i - y_m)^2 }$.
Note that computing Eq.~\eqref{eq:interpolation-neighbour} is more efficient than computing Eq.~\eqref{eq:interpolation}
because only the pixel values $I_m$ within the effective radius $\sigma$ needs to be selected, as depicted in Supplementary Fig.~\ref{fig:interpolation}.
The K-D~tree structure, generated earlier in the pixel registration step,
is thus utilized for fast searching of neighboring pixels of any given coordinates $(x_i,y_i)$.
The kernel $h(x,y)$, that also acts as a denoising filter, is currently approximated as a truncated 2D Gaussian function.
The quality of the restored image $\hat f(x,y)$ can be further improved by measuring the 2D point spread function (PSF) of the imaging setup
and the one dimensional impulse response of the antialiasing filter in the digitizer.

\begin{figure*}[p]
\centering\includegraphics[width=\textwidth]{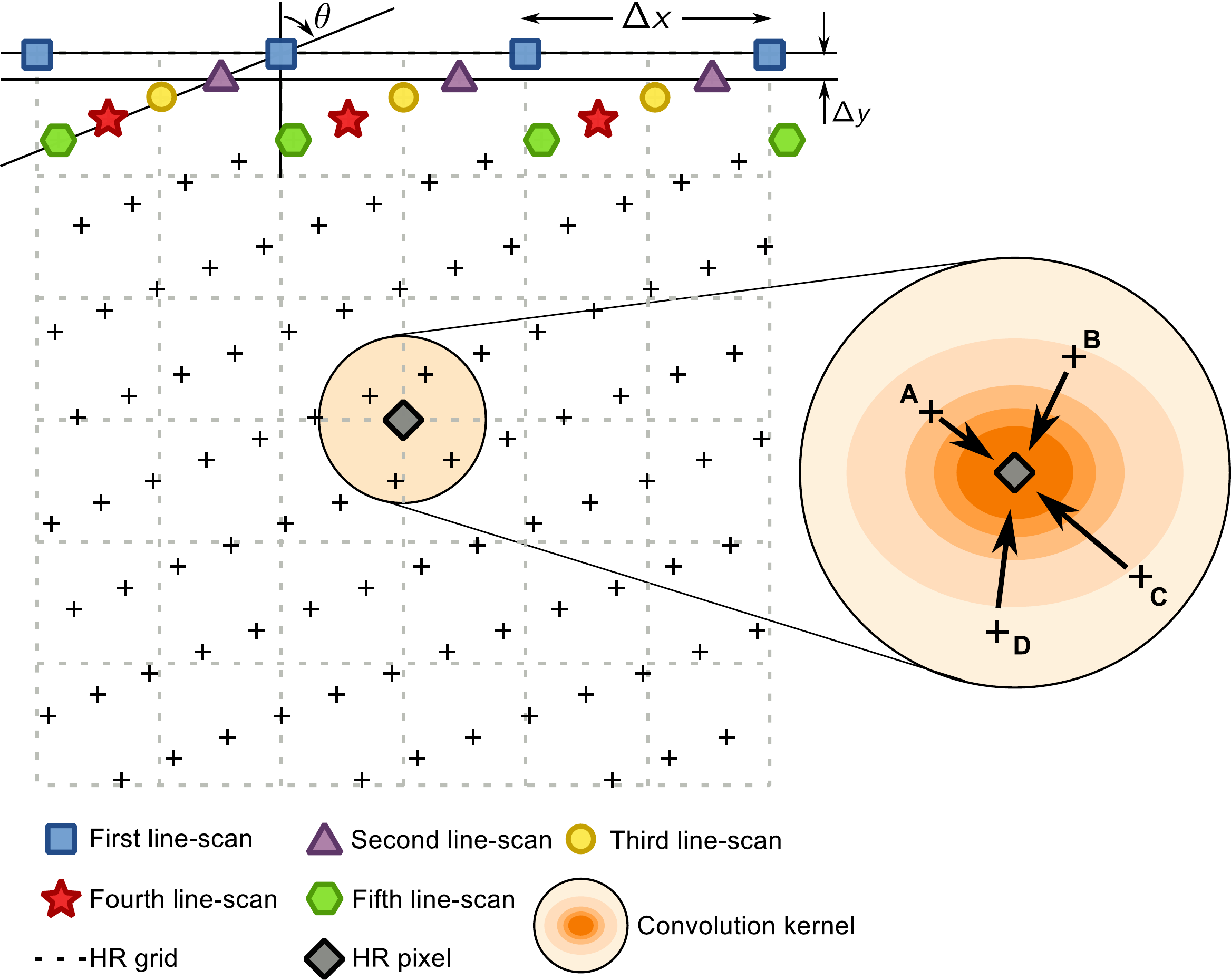}

\caption{\bf Illustration of the denoised non-uniform interpolation of the pixel-SR time-stretch image reconstruction algorithm.}
\label{fig:interpolation}
\end{figure*}

\end{document}